\documentclass[journal=nalefd,manuscript=letter]{achemso}

\usepackage[version=3]{mhchem} 
\usepackage[T1]{fontenc}       
\usepackage{color}

\newcommand{\RR}[0]{\boldsymbol{R}}
\newcommand{\kB}[0]{k_{\mathrm{B}}}
\newcommand{\ex}[1]{\mathrm{e}^{#1}}
\newcommand{\XX}[0]{\boldsymbol{X}}
\newcommand{\dd}[0]{\mathrm{d}}
\newcommand{\cc}[0]{\mathrm{c}}
\newcommand{\ee}[0]{\mathrm{e}}
\newcommand{\sss}[0]{\mathrm{s}}
\newcommand{\tot}[0]{\mathrm{tot}}
\author{Jaime Agudo-Canalejo}
\affiliation{Rudolf Peierls Centre for Theoretical Physics, University of Oxford, Oxford OX1 3NP, United Kingdom}
\alsoaffiliation{Department of Chemistry, The Pennsylvania State University, University Park, Pennsylvania 16802, United States}
\author{Pierre Illien}
\affiliation{Rudolf Peierls Centre for Theoretical Physics, University of Oxford, Oxford OX1 3NP, United Kingdom}
\alsoaffiliation{Department of Chemistry, The Pennsylvania State University, University Park, Pennsylvania 16802, United States}
\alsoaffiliation{ESPCI Paris, UMR Gulliver 7083, 10 rue Vauquelin, 75005 Paris, France}
\author{Ramin Golestanian}
\email{ramin.golestanian@ds.mpg.de, +495515176100}
\affiliation{Rudolf Peierls Centre for Theoretical Physics, University of Oxford, Oxford OX1 3NP, United Kingdom}
\alsoaffiliation{Max Planck Institute for Dynamics and Self-Organization (MPIDS), Am Fa{\ss}berg 17, D-37077 G\"{o}ttingen, Germany}

\title{Phoresis and Enhanced Diffusion Compete in Enzyme Chemotaxis}


\begin{document}

\begin{tocentry}

\includegraphics[width=0.58\linewidth]{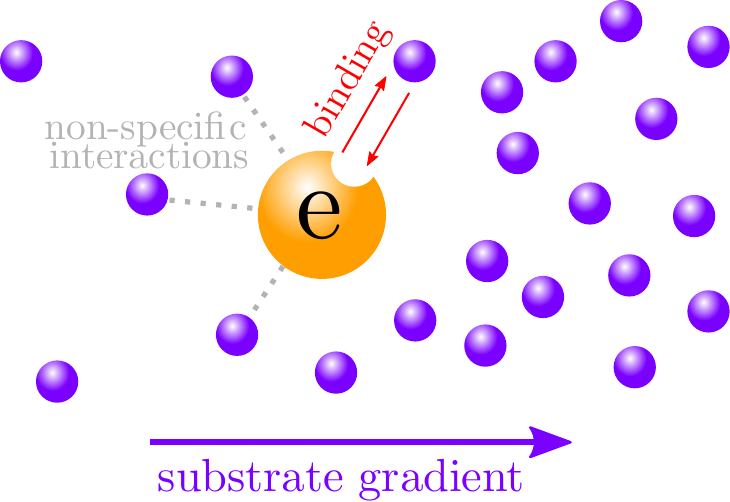}

\end{tocentry}

\begin{abstract}
Chemotaxis of enzymes in response to gradients in the concentration of their substrate has been widely reported in recent experiments, but a basic understanding of the process is still lacking. Here, we develop a microscopic theory for chemotaxis, valid for enzymes and other small molecules. Our theory includes both non-specific interactions between enzyme and substrate, as well as complex formation through specific binding between the enzyme and the substrate. We find that two distinct mechanisms contribute to enzyme chemotaxis: a diffusiophoretic mechanism due to the non-specific interactions, and a new type of mechanism due to binding-induced changes in the diffusion coefficient of the enzyme. The latter chemotactic mechanism points towards lower substrate concentration if the substrate enhances enzyme diffusion, and towards higher substrate concentration if the substrate inhibits enzyme diffusion. For a typical enzyme, attractive phoresis and binding-induced enhanced diffusion will compete against each other. We find that phoresis dominates above a critical substrate concentration, whereas binding-induced enhanced diffusion dominates for low substrate concentration. Our results resolve an apparent contradiction regarding the direction of urease chemotaxis observed in experiments, and in general clarify the relation between enhanced diffusion and chemotaxis of enzymes. Finally, we show that the competition between the two distinct chemotactic mechanisms may be used to engineer nanomachines that move towards or away from regions with a specific substrate concentration.
\end{abstract}


Because of their biocompatibility and ubiquity, enzymes have been extensively studied in recent years as ideal models for nanomachines.\cite{dey17} A subject of particular interest has been that of \emph{enhanced diffusion} of enzymes in the presence of their substrate: with increasing concentration of substrate, enhancements of up to $\sim 50\%$ in the diffusion coefficient of the enzyme have been measured for a wide variety of enzymes.\cite{yu09,mudd10,seng13,seng14,ried15,illi17a} Early hypotheses addressing enhanced diffusion of enzymes relied on non-equilibrium mechanisms such as stochastic swimming\cite{gole10,gole15,bai15} and exothermicity of the catalytic reaction.\cite{ried15,gole15,hwan17} However, it was observed recently that even the slow and endothermic enzyme aldolase undergoes enhanced diffusion. \cite{illi17a} In order to explain this observation, an equilibrium mechanism for the enhanced diffusion of enzymes has been proposed, which relies on the conformational changes of the enzyme due to binding and unbinding with the substrate. \cite{illi17a,illi17b}

More recently, a number of experimental studies have reported on the directed motion of enzymes in the presence of concentration gradients of their respective substrate, or \emph{chemotaxis}. The majority of studies so far have observed chemotaxis of the enzyme \emph{towards} their substrate: this includes RNA polymerase,\cite{yu09} catalase and urease, \cite{seng13,dey14} DNA polymerase, \cite{seng14} as well as hexokinase and aldolase. \cite{zhao17} Moreover, in all of these studies, enhanced diffusion of the enzyme in the presence of uniform substrate concentrations was reported, and it was hypothesized---albeit without a clear connection---that enhanced diffusion may be responsible for the observed chemotaxis.\cite{yu09,seng13,seng14} Recently, however, Jee \latin{et al.}\cite{jee17}~reported chemotaxis of urease and acetylcholinesterase \emph{away from} their respective substrate, while still observing enhanced diffusion for the two. The latter results appear to be in conflict with the previous observations for two reasons: first, the trend of enhanced diffusion being concomitant with chemotaxis towards the substrate is not followed; second, the observation of urease chemotaxis away from urea contradicts the earlier observation by Sengupta \latin{et al.}\cite{seng13}~of urease chemotaxis towards urea. In a somewhat different, non-enzymatic system, Guha \latin{et al.}\cite{guha17}~recently observed the chemotaxis of small molecular dyes towards higher concentrations of a large polymer to which they bind, while also observing \emph{inhibited diffusion} of the dye in the presence of the polymer. Whether there is a relation between chemotaxis and enhanced diffusion, and what exacty determines whether an enzyme or small molecule will chemotax towards or away from its substrate, are important questions that need to be addressed.

In this Letter, we present a microscopically detailed theory for enzyme chemotaxis. Our theory is a natural extension of our recently proposed equilibrium theory for the enhanced diffusion of enzymes,\cite{illi17a,illi17b} and does not rely on the catalytic activity (i.e. the non-equilibrium reaction step) of the enzyme. We take into account both non-specific interactions between enzyme and substrate, as well as complex formation through specific binding between the enzyme and the substrate. Here, and in the following, we use the term \emph{non-specific interactions} to refer to the forces to which all substrate molecules that are anywhere near the enzyme are subject to. These non-specific interactions can in principle be attractive or repulsive and include van der Waals, electrostatic, steric interactions, etc. On the other hand, we use the term \emph{specific binding} to refer to the attractive, short-ranged interactions that occur between a single substrate molecule and the enzyme at a well-defined binding pocket. These typically involve hydrogen bonding, hydrophobic interactions and temporary covalent bonds, and are often associated to a conformational change of the enzyme. We find that, for a typical enzyme, chemotaxis is dictated by a competition between two distinct mechanisms. The first one corresponds to diffusiophoresis, which arises from the non-specific interactions between enzyme and substrate, while the second one is due to the binding-induced changes in the diffusion coefficient of the enzyme. The contribution to chemotaxis due to binding-induced changes in diffusion points away from the substrate in the case of enhanced diffusion, and towards the substrate in the case of inhibited diffusion. We find a critical substrate concentration above which the diffusiophoretic mechanism dominates, and below which the mechanism arising from binding-induced changes in diffusion dominates. This competition can explain the conflicting experimental observations for urease, and is consistent with the rest of the available experimental evidence. Finally, we show that the competition between phoresis and binding-induced changes in diffusion can be used to engineer nanomachines that move towards or away from regions with a specific substrate concentration.

\begin{figure}
\begin{center}
\includegraphics[width=0.7\linewidth]{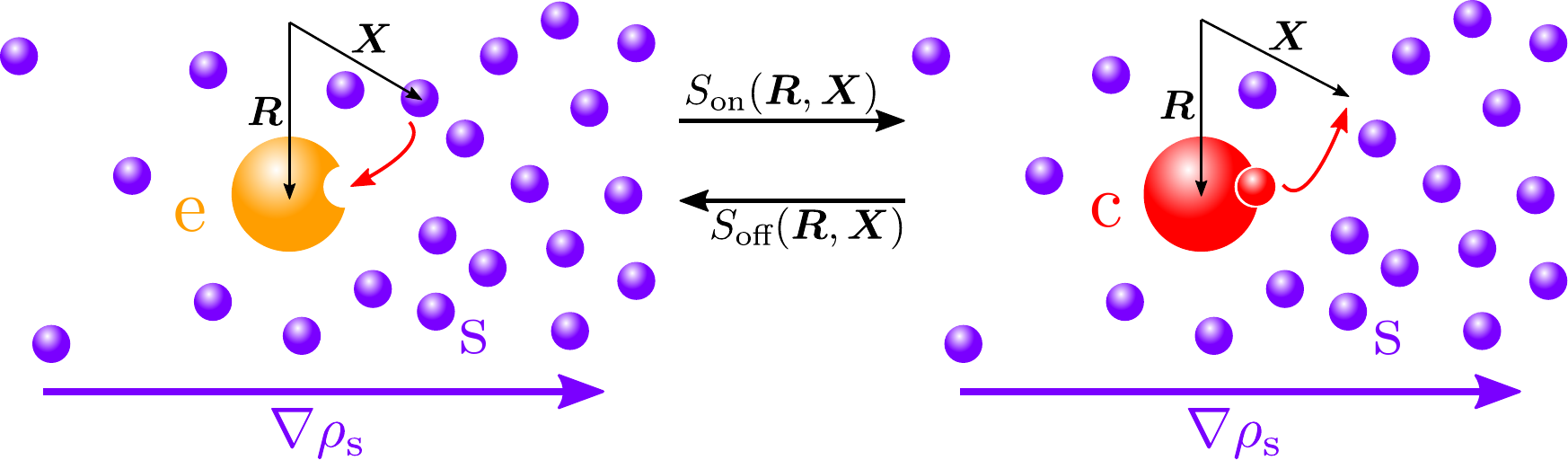}
\caption{A free enzyme (yellow) in a gradient $\nabla \rho_{\sss}$ of substrate molecules (purple). An enzyme--substrate complex (red) can form with probability $S_\text{on}(\RR,\XX)$ and decompose with probability $S_\text{off}(\RR,\XX)$.}
\label{fig:scheme}
\end{center}
\end{figure}

We begin by considering a single enzyme in a bath of $N$ substrate molecules, as depicted in Fig~\ref{fig:scheme}. The enzyme is located at $\RR$, while the substrate molecules are at $\XX_1,...,\XX_N$. An enzyme located at $\RR$ can bind to a substrate molecule located at $\XX$ with probability $S_\text{on}(\RR,\XX)$, to form a complex located at $\RR$.\cite{lee87} In turn, the complex located at $\RR$ can decompose, leaving the free enzyme at $\RR$ and the substrate molecule at $\XX$, with probability $S_\text{off}(\RR,\XX)$. The state where the enzyme is free is described by the $(N+1)$-particle probability density $\rho_\text{f}(\RR,\XX_1,...,\XX_N;t)$. There are $N$ distinct states where a complex coexists with $N-1$ free substrate molecules, each of these states being described by an $N$-particle probability density density $\rho_{\text{b},i}(\RR,\XX_1,...,\XX_{i-1},\RR,\XX_{i+1},...,\XX_N;t)$ for $1\leq i \leq N$. The non-specific interactions between the enzyme and the substrate molecules are taken into account \latin{via} an interaction potential $\phi^{\ee}(\RR,\XX_1,...,\XX_N)$ in the free state, and  $\phi^{\cc,i}(\RR,\XX_1,...,\XX_{i-1},\RR,\XX_{i+1},...,\XX_N)$ in the bound state. All particles may also interact with each other through hydrodynamic interactions. We can then write the following coupled set of Smoluchowski equations obeyed by the probability distributions characterising each of the $N+1$ distinct states
\begin{eqnarray}
&& \partial_t\rho_\text{f}(\RR,\XX_1,...,\XX_N)=\nonumber\\
&&  \nabla_{\RR} \cdot \boldsymbol{\mu}^{\ee \ee}\cdot[\kB T \nabla_{\RR}\rho_\text{f}+  (\nabla_{\RR}\phi^{\ee})\rho_\text{f}] +\sum_{i=1}^N \left\{\nabla_{\RR} \cdot \boldsymbol{\mu}^{\ee \sss}\cdot[\kB T \nabla_{\XX_i}\rho_\text{f}+ (\nabla_{\XX_i}\phi^{\ee})\rho_\text{f}] \right. \nonumber\\
&&+ \nabla_{\XX_i} \cdot \boldsymbol{\mu}^{\sss \ee}\cdot[\kB T \nabla_{\RR}\rho_\text{f}+  (\nabla_{\RR}\phi^{\ee})\rho_\text{f}]+\nabla_{\XX_i} \cdot \boldsymbol{\mu}^{\sss \sss}\cdot[\kB T \nabla_{\XX_i}\rho_\text{f}+ (\nabla_{\XX_i}\phi^{\ee})\rho_\text{f}]\big\} \label{eq:FP1} \\
&&-\sum_{i=1}^N S_\text{f}(\RR,\XX_i)\rho_\text{f}(\RR,\XX_1,...,\XX_N)+\sum_{i=1}^N S_\text{r}(\RR,\XX_i) \rho_{\text{b},i}(\RR,\XX_1,...,\XX_{i-1},\RR,\XX_{i+1},...,\XX_N) \nonumber
\end{eqnarray}
\begin{eqnarray}
&& \partial_t\rho_{\text{b},i}(\RR,\XX_1,...,\XX_{i-1},\RR,\XX_{i+1},...,\XX_N)=\nonumber\\
&&  \nabla_{\RR} \cdot \boldsymbol{\mu}^{\cc \cc}\cdot[\kB T \nabla_{\RR}\rho_{\text{b},i}  +  (\nabla_{\RR}\phi^{\cc,i})\rho_{\text{b},i}] +\sum_{\substack{j=1 \\  (j \neq i) }}^N \left\{\nabla_{\RR} \cdot \boldsymbol{\mu}^{\cc \sss}\cdot[\kB T \nabla_{\XX_j}\rho_{\text{b},i}+ (\nabla_{\XX_j}\phi^{\cc,j})\rho_{\text{b},i}] \right. \nonumber\\
&&+ \nabla_{\XX_j} \cdot \boldsymbol{\mu}^{\sss \cc}\cdot[\kB T \nabla_{\RR}\rho_{\text{b},i}+  (\nabla_{\RR}\phi^{\cc,j})\rho_{\text{b},i}]+\nabla_{\XX_j} \cdot \boldsymbol{\mu}^{\sss \sss}\cdot[\kB T \nabla_{\XX_j}\rho_{\text{b},i}+ (\nabla_{\XX_j}\phi^{\cc,j})\rho_{\text{b},i}]\big\} \label{eq:FP2} \\
&& + \int \dd \XX_i  S_\text{f}(\RR,\XX_i) \rho_\text{f}(\RR,\XX_1,...,\XX_N) - \rho_{\text{b},i}(\RR,\XX_1,...,\XX_{i-1},\RR,\XX_{i+1},...,\XX_N) \int \dd \XX_i  S_\text{r}(\RR,\XX_i) \nonumber
\end{eqnarray}
with $1 \leq i \leq N$, and $\boldsymbol{\mu}^{\alpha \beta}$ corresponding to the hydrodynamic mobility tensors.

We now seek to obtain evolution equations for the one-particle probability densities $\rho_{\ee}(\RR)$ and $\rho_{\cc}(\RR)$ representing the free enzyme and the complex. These result from integrating out the positional degrees of freedom corresponding to the substrate molecules, such that 
\begin{eqnarray}
\rho_{\ee}(\RR) & = & \int \dd \XX_1 ... \dd \XX_N\,  \rho_{\text{f}}(\RR,\XX_1,...,\XX_N), \\
 \rho_{\cc}(\RR) & = &  \sum_{i=1}^N \rho_{\cc,i}(\RR) ,~\text{with}\\
\rho_{\cc,i}(\RR) & = & \int \dd \XX_1 ... \dd \XX_{i-1} \dd\XX_{i+1} ... \dd \XX_N\,  \rho_{\text{b},i}(\RR,\XX_1,...,\XX_{i-1},\RR,\XX_{i+1},...,\XX_N).
\end{eqnarray}
In general, the mobility tensors $\boldsymbol{\mu}^{\alpha \beta}$ as well as the interaction potentials $\phi^{\alpha}$ may depend on the coordinates $\RR,\XX_1,...,\XX_N$ of all particles. We will, however, focus on the case of sufficiently dilute solutions and small substrate molecules, so that the self-mobilities $\boldsymbol{\mu}^{ii}$ can be considered to be constant, and the cross-mobilities $\boldsymbol{\mu}^{ij}$ (with $i \neq j$) include only the pair contribution. The interaction potentials are a sum of pairwise interactions, with $\phi^{\ee}=\sum_{i=1}^N \phi^{\ee \sss}(\RR,\XX_i)$ and $\phi^{\cc,i}=\sum_{\substack{j=1 \\ (j \neq i)}}^N \phi^{\cc \sss}(\RR,\XX_j)$. For the simplest description of binding and unbinding, we may use $S_\text{on}(\RR,\XX) = k_\mathrm{on} \delta(\RR-\XX)$ and $S_\text{off}(\RR,\XX) = k_\mathrm{off} \delta(\RR-\XX)$, representing the fact that binding and unbinding are due to very short-ranged interactions and occur with rates $k_\mathrm{on}$ and $k_\mathrm{off}$.

Using these approximations, we can integrate Equations~(\ref{eq:FP1}--\ref{eq:FP2}) to obtain the two coupled equations for the free enzyme and complex probability distributions (see Supporting Information for details of the derivation)
\begin{eqnarray}
\partial_t \rho_{\ee}(\RR;t) & = &  \nabla_{\RR}\cdot \left[  D_{\ee} \nabla_{\RR}\rho_{\ee}  -  \boldsymbol{v_{\ee}}(\RR) \rho_{\ee} \right]  - k_\mathrm{on} \rho_{\ee} \rho_{\sss} + k_\mathrm{off} \rho_{\cc} \label{eq:coupled1} \\
\partial_t \rho_{\cc}(\RR;t) & = &  \nabla_{\RR}\cdot \left[  D_{\cc} \nabla_{\RR}\rho_{\cc} -   \boldsymbol{v_{\cc}}(\RR) \rho_{\cc} \right]  + k_\mathrm{on} \rho_{\ee} \rho_{\sss} - k_\mathrm{off} \rho_{\cc} \label{eq:coupled2}
\end{eqnarray}
where $\rho_{\sss}(\RR)$ is the concentration of substrate molecules. Equations (\ref{eq:coupled1}--\ref{eq:coupled2}) capture the effects of three important physical mechanisms. First, the free enzyme can turn into a complex and \latin{vice versa} through the \emph{binding and unbinding} of a substrate molecule, with rates $k_\mathrm{on}$ and $k_\mathrm{off}$. Second, the free enzyme and complex \emph{diffuse} with diffusion coefficients respectively given by $D_{\ee} = \kB T \boldsymbol{\mu}^{\ee \ee}$ and $D_{\cc} = \kB T \boldsymbol{\mu}^{\cc \cc}$. Finally, the combination of non-specific and hydrodynamic interactions between the free enzyme or complex and the substrate molecules leads to a \emph{diffusiophoretic drift} of the free enzyme and complex with velocities respectively given by
\begin{eqnarray}
\boldsymbol{v_{\ee}}(\RR) & \approx & \frac{\kB T}{\eta} \left[ \int_0^\infty \mathrm{d}h h \left( \ex{- \phi^{\ee \sss}(h)/\kB T} -1 \right) \right] \nabla_{\RR} \rho_{\sss} \equiv \frac{\kB T}{\eta} \lambda_{\ee}^2 \nabla_{\RR} \rho_{\sss} \label{eq:v1}  \\
\boldsymbol{v_{\cc}}(\RR) & \approx & \frac{\kB T}{\eta} \left[ \int_0^\infty \mathrm{d}h h \left( \ex{- \phi^{\cc \sss}(h)/\kB T} -1 \right) \right] \nabla_{\RR} \rho_{\sss} \equiv \frac{\kB T}{\eta} \lambda_{\cc}^2 \nabla_{\RR} \rho_{\sss} \label{eq:v2}
\end{eqnarray}
where $\eta$ is the viscosity of the fluid,  and $h$ is the distance between the surface of the enzyme and the center of the substrate molecule. Equations (\ref{eq:v1}--\ref{eq:v2}) are approximate forms of the diffusiophoretic velocity valid for the typical case in which the range of the non-specific interactions is shorter than the size of the enzyme/complex (in the general case, there are corrections to these velocities, see Supporting Information). We have defined here the \emph{Derjaguin length} $\lambda_\alpha$, a material parameter that captures the non-specific interactions between the enzyme or complex and the substrate.\cite{derj47,ebbe12} Typical values of the Derjaguin length are of a few angstroms.\cite{ebbe12,ande89} We note that, in this convention, $\lambda_\alpha^2$ may be positive or negative, with positive (negative) values corresponding to overall attractive (repulsive) interactions that lead to an enrichment (depletion) of substrate molecules in the proximity of the enzyme. Non-specific interactions that are always present for every enzyme--substrate pair include repulsive steric interactions, and attractive van der Waals interactions. In the Supporting Information, we show that even rather weak and short-ranged attractive interactions are sufficient to compensate for steric repulsion and induce a positive $\lambda_\alpha^2>0$ for a typical enzyme, in which case the phoretic velocity will be directed towards higher concentrations of the substrate.

We note that, in the mechanism described above, we have neglected the catalytic step of the enzymatic reaction by which substrate molecules can be turned into product molecules. This catalytic step has two main consequences. First, it represents an alternative pathway by which a complex can dissociate and give back a free enzyme. 
However, since the rate $k_\mathrm{cat}$ of the catalytic step is typically much slower than the unbinding rate ($k_\mathrm{cat} \ll k_\mathrm{off}$) the effect of the catalytic step is therefore often negligible in this regard. Second, when catalysis is taken into account the number of substrate molecules is no longer conserved, and decreases in time. Nevertheless, as long as the concentration of enzyme in solution is much lower than the concentration of substrate, and the experiments are performed in a sufficiently short time, the decrease in substrate concentration due to catalysis may be neglected.

Equations (\ref{eq:coupled1}--\ref{eq:coupled2}) already contain all the ingredients necessary to describe enzyme--substrate interactions. Nevertheless, in order to describe chemotaxis we are actually interested in the \emph{total} concentration of enzyme, both free and bound, given by
\begin{equation}
\rho_{\ee}^{\tot}(\RR;t) = \rho_{\ee}(\RR;t) + \rho_{\cc}(\RR;t) \label{eq:rhotot}
\end{equation}
which corresponds to what is actually measured in experiments with fluorescently tagged enzymes (free enzyme and complex cannot be distinguished). Furthermore, the typical timescale of diffusion and phoretic drift is much longer than the typical timescale of binding and unbinding. We can therefore assume that the enzyme is locally and instantaneously at binding equilibrium with the substrate, so that at any position $\RR$ we will have
\begin{equation}
k_\mathrm{on} \rho_{\ee}(\RR;t) \rho_{\sss}(\RR;t) \approx k_\mathrm{off} \rho_{\cc}(\RR;t) \label{eq:fasteq}
\end{equation}
at time $t$. Combining (\ref{eq:rhotot}) and (\ref{eq:fasteq}), we find the typical Michaelis-Menten kinetics for the free enzyme and the complex
\begin{equation}
\rho_{\ee} = \frac{K}{K + \rho_{\sss}}\rho_{\ee}^{\tot}~~\text{and}~~\rho_{\cc} = \frac{\rho_{\sss}}{K + \rho_{\sss}}\rho_{\ee}^{\tot} \label{eq:MichM}
\end{equation}
where we have defined the Michaelis constant $K \equiv k_\mathrm{off}/k_\mathrm{on}$.

Adding together Equations (\ref{eq:coupled1}) and (\ref{eq:coupled2}), and using (\ref{eq:MichM}), we finally obtain an expression for the time evolution of the total enzyme concentration
\begin{equation}
\partial_t \rho_{\ee}^{\tot}(\RR;t) = \nabla_{\RR} \cdot \left\{ D(\RR)\cdot\nabla_{\RR}\rho_{\ee}^{\tot} - [\boldsymbol{V}_\mathrm{ph}(\RR) + \boldsymbol{V}_\text{bi}(\RR)] \rho_{\ee}^{\tot} \right\}
\label{eq:evol}
\end{equation}
with the space-dependent diffusion coefficient
\begin{eqnarray}
D(\RR) &=& D_{\ee} + (D_{\cc} - D_{\ee})   \frac{\rho_{\sss}(\RR)}{K + \rho_{\sss}(\RR)},
\end{eqnarray}
a drift velocity arising from phoretic effects
\begin{eqnarray}
\boldsymbol{V}_\mathrm{ph}(\RR) &=& \boldsymbol{v_{\ee}}(\RR) + [\boldsymbol{v_{\cc}}(\RR)-\boldsymbol{v_{\ee}}(\RR)]  \frac{\rho_{\sss}(\RR)}{K + \rho_{\sss}(\RR)},
\label{eq:Vph}
\end{eqnarray}
as well as a drift velocity arising from the changes in diffusion coefficient due to substrate binding and unbinding
\begin{eqnarray}
\boldsymbol{V}_\mathrm{bi}(\RR) &=& - (D_{\cc} - D_{\ee}) \nabla_{\RR} \left(  \frac{\rho_{\sss}(\RR)}{K + \rho_{\sss}(\RR)} \right).
\label{eq:Vbi}
\end{eqnarray}

Equation (\ref{eq:evol}) consitutes one of the main results of this work, and presents a number of interesting features. First, we notice that the diffusion coefficient $D(\RR)$ corresponds to the `average' of the diffusion coefficients of the free enzyme and the complex. In the absence of substrate $\rho_{\sss}=0$, the diffusion coefficient is that of the free enzyme. With increasing substrate concentration, the diffusion coefficient approaches that of the complex, saturating to $D_{\cc}$ for $\rho_{\sss} \gg K$. This Michaelis-Menten-like dependence of the diffusion coefficient of the enzyme with the substrate concentration has been observed experimentally, and is consistent with the equilibrium theory for enhanced diffusion introduced in Refs~\citenum{illi17a} and \citenum{illi17b}, which can explain the enhanced diffusion of slow, endothermic enzymes such as aldolase, \cite{illi17a} in contrast to other mechanisms proposed in the literature \cite{ried15,gole15,jee17} which rely on the exothermicity of the catalytic reaction. In Refs~\citenum{illi17a} and \citenum{illi17b}, several mechanisms where identified by which binding can induce changes in the diffusion coefficient of an enzyme, of the order of a few ten percent. These include binding-induced changes in the hydrodynamic radius of the enzyme, or changes in the fluctuations of the internal translational and rotational degrees of freedom of individual subunits of a modular enzyme. These changes were predicted to lead typically to $D_{\cc}>D_{\ee}$, corresponding to \emph{enhanced} diffusion. The case with $D_{\cc}<D_{\ee}$, on the other hand, would correspond to \emph{inhibited} diffusion. At any rate, the theory presented here is independent of the underlying mechanism that may cause $D_{\ee}$ and $D_{\cc}$ to be different from each other, and only depends on their actual values. In practice, the values of $D_{\ee}$ and $D_{\cc}$ can be extracted from the available experimental data, as they correspond to the measured diffusion coefficients of the enzyme in the absence of substrate and in saturating substrate concentration, respectively.

The drift velocity arising from diffusiophoresis $\boldsymbol{V}_\mathrm{ph}(\RR)$ also corresponds to an `average' of the phoretic velocities of the free enzyme and the complex. With increasing substrate concentration, we again find a smooth Michaelis-Menten-like crossover between the velocity of the free enzyme $\boldsymbol{v_{\ee}}$ and that of the complex $\boldsymbol{v_{\cc}}$. In principle, this velocity may be directed towards or away from higher concentrations of substrate, depending on the details of the non-specific interactions. Nevertheless, as mentioned previously, because enzyme--substrate interactions are generally attractive we expect that the typical phoretic velocity for enzymes will be directed towards higher substrate concentrations.

Finally, the drift velocity $\boldsymbol{V}_\mathrm{bi}(\RR)$ is a direct consequence of the changes in the diffusion coefficient of the enzyme due to binding and unbinding of the substrate. This drift velocity is directed towards higher concentrations of substrate in the case of inhibited diffusion with $D_{\cc}<D_{\ee}$, and towards lower concentrations of substrate in the case of enhanced diffusion with $D_{\cc}>D_{\ee}$. This tendency can be made even more apparent when we notice that $\boldsymbol{V}_\mathrm{bi}(\RR)$ can alternatively be written as $\boldsymbol{V}_\mathrm{bi}(\RR)= - \nabla_{\RR} D(\RR)$. In the absence of phoresis with $\boldsymbol{V}_\mathrm{ph}(\RR)=0$, Equation (\ref{eq:evol}) can then be written as $\partial_t \rho_{\ee}^{\tot}(\RR;t) = \nabla_{\RR}^2[D(\RR) \rho_{\ee}^{\tot}]$, and we would thus expect $\rho_{\ee}^{\tot}(\RR) \propto 1/D(\RR)$ in the steady state, i.e.~the enzyme tends to concentrate in regions where its diffusion is slowest. This type of behaviour was recently reported experimentally in Ref~\citenum{jee17} for urease, and was explored theoretically in Ref~\citenum{weis17}. However, in these two works the equation $\partial_t \rho_{\ee}^{\tot}(\RR;t) = \nabla_{\RR}^2[D(\RR) \rho_{\ee}^{\tot}]$ was simply postulated with no justification, even if it is an established fact that knowledge of the position-dependence of the diffusion coefficient is not sufficient to obtain an evolution equation for the system: due to what is known as \emph{multiplicative noise}, different evolution equations, such as e.g.~$\partial_t \rho_{\ee}^{\tot}(\RR;t) = \nabla_{\RR} [D(\RR) \nabla_{\RR} \rho_{\ee}^{\tot}]$, may arise from different microscopic mechanisms underlying the position-dependence of the diffusion coefficient.\cite{schn93,lau07} Here, we have provided for the first time a microscopic mechanism that can lead to a steady state with $\rho_{\ee}^{\tot}(\RR) \propto 1/D(\RR)$.

The results from Ref~\citenum{jee17} just described, in which urease was observed to chemotax \emph{away from} higher concentrations of urea, are in apparent conflict with older results in the literature, Ref~\citenum{seng13}, in which urease was observed to chemotax \emph{towards} higher concentrations of urea. The existence of two distinct mechanisms for chemotaxis, namely phoresis and binding-induced changes in diffusion as just described, may explain the seemingly contradictory observations. For simplicity, let us consider the typical case in which the phoretic response of the free enzyme and the complex is similar, with $\lambda_{\ee} = \lambda_{\cc}$. This is expected if the binding-unbinding process does not significantly affect the surface chemistry of the enzyme, so that the non-specific interactions with the substrate are identical for the free enzyme and the complex. In this case, the phoresis-induced velocity (\ref{eq:Vph}) is simply
\begin{eqnarray}
\boldsymbol{V}_\mathrm{ph} &=& \frac{\kB T}{\eta} \lambda_{\ee}^2 \nabla_{\RR} \rho_{\sss} = 6 \pi D_{\ee} R_{\ee} \lambda_{\ee}^2 \nabla_{\RR} \rho_{\sss}
\label{eq:Vph2}
\end{eqnarray}
where we have used the Stokes-Einstein relation to express $\kB T/\eta$ in terms of the diffusion coefficient $D_{\ee}$ and hydrodynamic radius $R_{\ee}$ of the enzyme. In turn, the velocity (\ref{eq:Vbi}) due to binding-induced changes in diffusion can be written equivalently as
\begin{eqnarray}
\boldsymbol{V}_\mathrm{bi} &=& - \alpha D_{\ee} \frac{K}{(K + \rho_{\sss})^2} \nabla_{\RR} \rho_{\sss}
\label{eq:Vbi2}
\end{eqnarray}
where $\alpha \equiv (D_{\cc} - D_{\ee})/D_{\ee}$ represents the dimensionless change of diffusion coefficient between the free enzyme and the complex, which is positive and negative for enhanced and inhibited diffusion, respectively. Typical values observed experimentally for enhanced diffusion correspond to $\alpha = 0.1-0.5$. Now, we observe that while both the phoretic and the binding-induced velocity are proportional to the substrate concentration gradient $\nabla_{\RR} \rho_{\sss}$, the former is independent of the actual substrate concentration $\rho_{\sss}$, whereas the magnitude of the latter decreases with increasing substrate concentration. Both contributions will have the same magnitude whenever $|\boldsymbol{V}_\mathrm{ph}| = |\boldsymbol{V}_\mathrm{bi}|$, or equivalently when
\begin{eqnarray}
\rho_{\sss} = K \left( \sqrt{ \frac{|\alpha|}{6 \pi R_{\ee} |\lambda_{\ee}^2| K} } - 1 \right) \equiv \rho_{\sss}^*
\label{eq:critical}
\end{eqnarray}
which defines a \emph{critical substrate concentration} $\rho_{\sss}^*$, above which phoresis dominates over binding-induced changes in diffusion, and below which the latter dominate over the former. For any given enzyme--substrate pair with their corresponding enzyme radius $R_{\ee}$, Derjaguin length $\lambda_{\ee}$, Michaelis constant $K$, and diffusion change $\alpha$, there will be a different critical substrate concentration. We also note that phoresis may dominate at all concentrations if the binding-induced changes in diffusion are too small, with $|\alpha| \leq 6 \pi R_{\ee} |\lambda_{\ee}^2| K$.

In the urease experiments of Jee \latin{et al.},\cite{jee17} in which chemotaxis away from urea was observed, the urea concentration used was 1~mM. In those of Sengupta \latin{et al.},\cite{seng13} in which chemotaxis towards urea was observed, the urea concentration was 1~M. It is therefore possible that, if the critical substrate concentration for this system lies between these two values, the former experiments are dominated by the binding mechanism, which necessarily points away from urea because urease shows enhanced diffusion with $\alpha>0$, whereas the latter experiments are dominated by phoresis, which may be directed towards urea as long as $\lambda_{\ee}^2>0$, which is expected for typical enzymes. Let us now test the plausibility of the proposed hypothesis. The hydrodynamic radius of urease is $R_{\ee}=7$~nm,\cite{mudd10} the Michaelis constant of urea--urease is $K=3$~mM, \cite{kraj09} and the diffusion enhancement is $\alpha=0.3$. \cite{mudd10} Using these values and Equation (\ref{eq:critical}), we calculate that the critical substrate concentration $\rho_{\sss}^*$ will be between 1~M and 1~mM as long as the Derjaguin length $\lambda_{\ee}$ of urease--urea is between 0.034\;\AA~and 8.4\;\AA, precisely within the typical range expected for a Derjaguin length. \cite{ebbe12,ande89} In practice, for the Jee \latin{et al.}~and the Sengupta \latin{et al.}~experiments to be well within the regimes of dominance of the binding and the phoresis mechanisms, respectively, we would expect the Derjaguin length to be somewhere far enough from the two extreme values of the range, say, between 0.1\;\AA~and 3\;\AA. As an example, for $\lambda_{\ee}=1$\;\AA~we find the critical urea concentration $ \rho_{\sss}^* \simeq 30$~mM, which is simultaneously much larger than 1~mM and much smaller than 1~M. As described in the Supporting Information, even rather weak ($\sim 1~\kB T$) and short-ranged ($\sim 0.75$~\AA) attractive interactions, consistent with e.g.~van der Waals forces, are sufficient to induce attractive phoresis with $\lambda_{\ee}=1$\;\AA.

The same type of consistency check can be applied to other studies of chemotaxis of enzymes that show enhanced diffusion. In Ref~\citenum{seng13}, catalase was observed to chemotax towards its substrate, hydrogen peroxide, so we would expect the experiment to be dominated by phoresis. Using the radius $R_\mathrm{e}=4$~nm of catalase,\cite{seng13} the diffusion enhancement observed $\alpha=0.45$,\cite{seng13} the Michaelis constant $K=93$~mM,\cite{swit02} and the substrate concentration used in the experiments $\rho_{\sss}=10$~mM,\cite{seng13} we find that those experiments could be dominated by phoresis if $\lambda_{\ee}>2.9$\;\AA, a somewhat high value but still within plausible range. In Ref~\citenum{zhao17}, hexokinase was observed to chemotax towards its substrate, D-glucose, again implying dominance of phoresis. Using the corresponding values\cite{zhao17} $R_\mathrm{e}=3.5$~nm, $\alpha=0.38$, $K=0.12$~mM, and $\rho_{\sss}=10$~mM, we expect the experiments to be dominated by phoresis if $\lambda_{\ee}>1$\;\AA, well within reasonable range. In Ref~\citenum{jee17}, acetylcholinesterase was observed to chemotax away from acetylcholine, implying dominance of binding-induced enhanced diffusion in this case. Indeed, using the corresponding values\cite{jee17} $R_\mathrm{e}=11$~nm, $\alpha=0.2$, $K=0.2$~mM,\cite{hu06,poha11} and $\rho_{\sss}=0.2$~mM, we find that phoresis could dominate only if $\lambda_{\ee}>14$\;\AA, which is an unreasonably high value.

For completeness, it is worth noting that because the theory presented above does not depend on catalysis in any way, it may be applied not only to enzymes, but also to the chemotaxis of small molecules that bind to other molecules. In Ref~\citenum{guha17}, chemotaxis of the molecular dye Rhodamine 6G towards higher concentrations of the polymer Ficoll 400K was observed. Unfortunately, in this particular system our theoretical description of diffusiophoresis is not expected to apply, given that the polymer (which represents the `substrate') is much larger than the dye (the `enzyme'), in violation of the assumptions of our derivation (see Supporting Information). The chemotactic contribution of binding-induced changes in diffusion, however, should still be applicable. Indeed, in this case binding of the dye to the polymer is expected to inhibit the dye diffusion: because the large polymer diffuses much more slowly than the small dye, we can expect the diffusion coefficient of the complex $D_{\cc}$ to be that of the polymer, i.e.~$D_{\cc} \approx D_\mathrm{pol} \ll D_\mathrm{dye}$. The change in diffusion coefficient is thus negative with $\alpha \approx (D_\mathrm{pol} - D_\mathrm{dye})/D_\mathrm{dye} \approx -1$, and the binding-induced velocity (\ref{eq:Vbi2}) therefore induces chemotaxis towards higher concentrations of the polymer, as observed experimentally.

In the discussion above, we focused for simplicity on the (presumably most common) case in which the non-specific interactions with the substrate are similar for the free enzyme and the complex, leading to similar values of the Derjaguin length $\lambda_\mathrm{e} \approx \lambda_\mathrm{c}$ for the two. Generally, however, $\lambda_\mathrm{e}$ and $\lambda_\mathrm{c}$ may be different from each other: in this case, as described by our full expression (\ref{eq:Vph}), the magnitude of the phoretic velocity is affected by substrate binding, and as a consequence also depends on substrate concentration. Nevertheless, even in this general case, it is is easy to show that there is a critical substrate concentration above and below which phoresis and changes in diffusion dominate, respectively. Two particularly interesting limiting cases correspond to those in which the non-specific interactions with the substrate are negligible either in the free or in the bound state of the enzyme, as would occur if the binding-induced conformational changes of the enzyme significantly affect its surface properties (e.g.~surface charges). If the free enzyme is non-interacting ($\lambda_{\ee}=0$ and $\lambda_{\cc} \neq 0$), the phoretic velocity is given by $\boldsymbol{V}_\mathrm{ph} =  \frac{\kB T}{\eta} \lambda_{\cc}^2 \frac{\rho_{\sss}}{K + \rho_{\sss}} \nabla_{\RR} \rho_{\sss}$, and increases in magnitude with increasing substrate concentration. The critical substrate concentration above which phoresis dominates is then given by $\rho_{\sss}^* \equiv  \frac{K}{2} \left( \sqrt{1 + \frac{2|\alpha|}{3 \pi R_{\ee} |\lambda_{\cc}^2| K} } - 1 \right)$. On the other hand, if the complex is non-interacting ($\lambda_{\ee} \neq 0$ and $\lambda_{\cc} = 0$), the phoretic velocity is given by $\boldsymbol{V}_\mathrm{ph} =  \frac{\kB T}{\eta} \lambda_{\ee}^2 \frac{K}{K + \rho_{\sss}} \nabla_{\RR} \rho_{\sss}$, and decreases in magnitude with increasing substrate concentration, although this decrease is less pronounced than that of the velocity due to binding-induced changes in the diffusion coefficient (\ref{eq:Vbi2}). For this reason, there will still be a critical substrate concentration above which phoresis dominates, now given by $\rho_{\sss}^* \equiv K \left( \frac{|\alpha|}{6 \pi R_{\ee} |\lambda_{\ee}^2| K} - 1 \right)$.

\begin{figure}
\begin{center}
\includegraphics[width=0.7\linewidth]{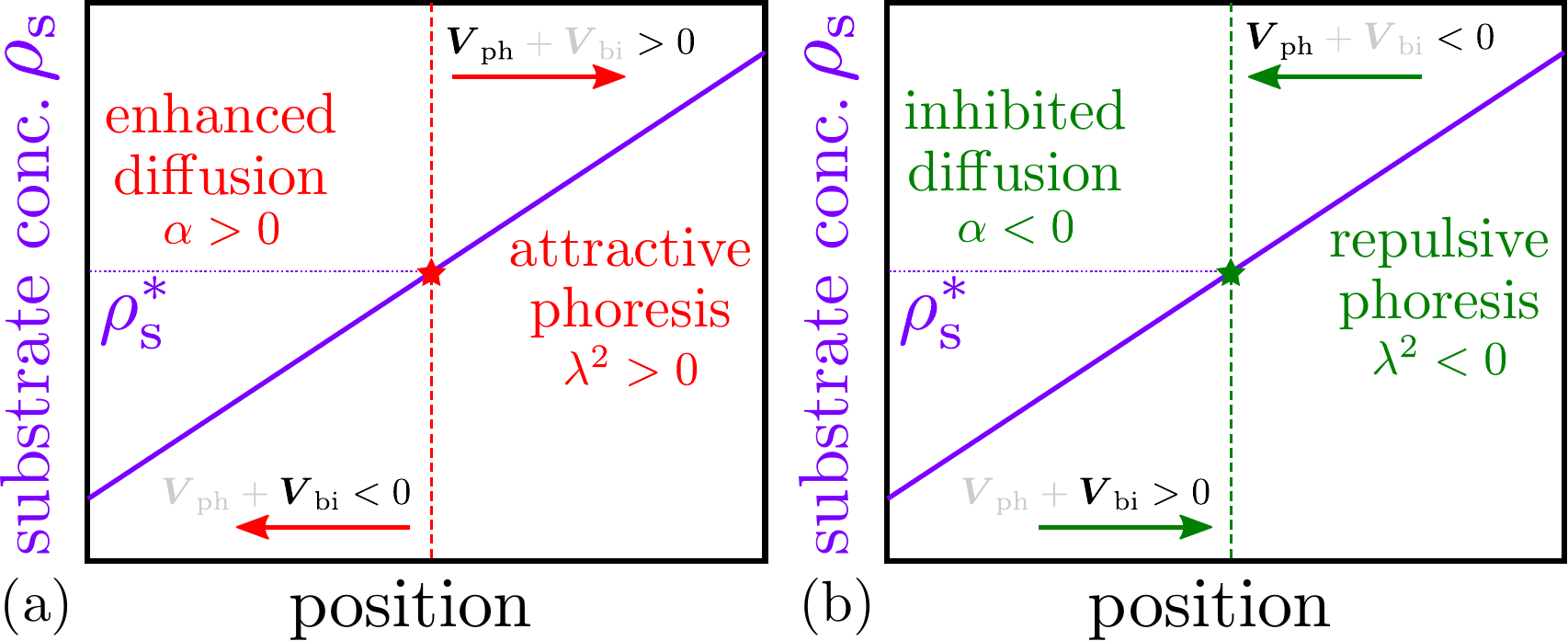}
\caption{Competition between phoresis and binding-induced changes in diffusion in an externally imposed substrate gradient. (a) A particle subject to binding-induced enhanced diffusion and attractive phoresis will be repelled from regions where the substrate concentration is equal to the critical concentration $\rho_{\sss}^*$. (b) A particle subject to binding-induced inhibited diffusion and repulsive phoresis will be attracted to regions where the substrate concentration is equal to the critical concentration $\rho_{\sss}^*$.}
\label{fig:gradients}
\end{center}
\end{figure}

The competition between phoresis and binding-induced changes in diffusion described in this Letter has some general consequences that may be useful in the design and engineering of intelligent nanomachines. When discussing enzymes, we have focused on the case of enhanced diffusion ($\alpha>0$) competing against attractive diffusiophoresis ($\lambda^2>0$). In this case, for concentrations of substrate above the critical value $\rho_{\sss}^*$, phoresis will dominate and the nanomachine will drift towards higher concentrations of the substrate. For concentrations below the critical value, binding-induced enhanced diffusion will dominate and the nanomachine will drift towards lower concentrations of the substrate. This implies that, in a steady-state gradient of substrate, the nanomachine will effectively be repelled from regions with the critical substrate concentration $\rho_{\sss}^*$, see Fig~\ref{fig:gradients}(a). Alternatively, it may be possible to engineer nanomachines for which binding of the substrate inhibits diffusion ($\alpha<0$), while non-specific interactions with the substrate lead to repulsive diffusiophoresis ($\lambda^2<0$). Now, for $\rho_{\sss}>\rho_{\sss}^*$, phoresis will dominate and the nanomachine will drift towards lower concentrations of the substrate, whereas for $\rho_{\sss}<\rho_{\sss}^*$, binding-induced inhibited diffusion will dominate and the nanomachine will drift towards higher substrate concentrations. Therefore, in this case the nanomachine will be effectively attracted to regions with the critical substrate concentration $\rho_{\sss}^*$, see Fig~\ref{fig:gradients}(b). Exploiting the competition between phoresis and binding-induced changes in diffusion is therefore a promising avenue to achieve finely-tuned self-organization at the nanoscale. However, we note that such complex behavior is not always expected to occur: in systems that display enhanced diffusion ($\alpha>0$) and repulsive diffusiophoresis ($\lambda^2<0$), or alternatively inhibited diffusion ($\alpha<0$) and attractive diffusiophoresis ($\lambda^2>0$), the two mechanisms will collaborate instead of competing, and the direction of chemotaxis will not change with substrate concentration.

In summary, we have developed a microscopically detailed theory for the chemotaxis of enzymes (and other small molecules) in the presence of gradients of their substrate. We take into account both the non-specific interactions between enzyme and substrate, as well as complex formation through specific binding of the enzyme to the substrate. The experimentally observed Michaelis-Menten-like dependence of the enzyme diffusion coefficient on substrate concentration arises naturally in the theory, and we find a novel contribution to chemotaxis due to binding-induced changes in the diffusion coefficient of the enzyme, which points away from or towards higher substrate concentration depending on whether the substrate enhances or inhibits the diffusion of the enzyme, respectively. In typical cases, the binding-induced contribution to the chemotactic velocity will compete with the diffusiophoretic contribution that arises from the non-specific enzyme--substrate interactions. Because the two contributions depend differently on the substrate concentration, phoresis will dominate at high substrate concentration, whereas binding-induced changes in diffusion will dominate at low substrate concentration. In this way, we could resolve an apparent contradiction regarding the experimentally-observed direction of urease chemotaxis in the presence of urea. We have further checked that our theory is consistent with the available experimental evidence for other enzymes and small molecules. Moreover, the theory can be experimentally tested in a straightforward way by studying the chemotaxis of enzymes in the presence of varying concentrations of their substrate. The competition between phoresis and binding-induced changes in diffusion described here may be harnessed to engineer nanomachines that are directed towards or away from regions with a specific concentration of a substrate. As a closing remark, it is worth noting that, although we have studied here the effects of substrate concentration, the collective behaviour that may arise at high enzyme concentrations \cite{saha14,mikh15} remains to be explored, both experimentally and theoretically.

\begin{acknowledgement}

We thank A.~Sen and F.~Mohajerani for stimulating discussions and for a critical reading of the manuscript. All authors ackowledge support from the US National Science Foundation under MRSEC Grant number DMR-1420620.

\end{acknowledgement}

\begin{suppinfo}
Details of the derivation of the one-particle distribution functions; derivation of the diffusiophoretic velocities; estimating values of the Derjaguin length for attractive potentials. Ref~\citenum{feld77} is cited in the Supporting Information.
\end{suppinfo}

\bibliography{biblio}

\providecommand{\latin}[1]{#1}
\makeatletter
\providecommand{\doi}
  {\begingroup\let\do\@makeother\dospecials
  \catcode`\{=1 \catcode`\}=2 \doi@aux}
\providecommand{\doi@aux}[1]{\endgroup\texttt{#1}}
\makeatother
\providecommand*\mcitethebibliography{\thebibliography}
\csname @ifundefined\endcsname{endmcitethebibliography}
  {\let\endmcitethebibliography\endthebibliography}{}
\begin{mcitethebibliography}{31}
\providecommand*\natexlab[1]{#1}
\providecommand*\mciteSetBstSublistMode[1]{}
\providecommand*\mciteSetBstMaxWidthForm[2]{}
\providecommand*\mciteBstWouldAddEndPuncttrue
  {\def\EndOfBibitem{\unskip.}}
\providecommand*\mciteBstWouldAddEndPunctfalse
  {\let\EndOfBibitem\relax}
\providecommand*\mciteSetBstMidEndSepPunct[3]{}
\providecommand*\mciteSetBstSublistLabelBeginEnd[3]{}
\providecommand*\EndOfBibitem{}
\mciteSetBstSublistMode{f}
\mciteSetBstMaxWidthForm{subitem}{(\alph{mcitesubitemcount})}
\mciteSetBstSublistLabelBeginEnd
  {\mcitemaxwidthsubitemform\space}
  {\relax}
  {\relax}

\bibitem[Dey and Sen(2017)Dey, and Sen]{dey17}
Dey,~K.~K.; Sen,~A. {Chemically Propelled Molecules and Machines}. \emph{J. Am.
  Chem. Soc.} \textbf{2017}, \emph{139}, 7666--7676\relax
\mciteBstWouldAddEndPuncttrue
\mciteSetBstMidEndSepPunct{\mcitedefaultmidpunct}
{\mcitedefaultendpunct}{\mcitedefaultseppunct}\relax
\EndOfBibitem
\bibitem[Yu \latin{et~al.}(2009)Yu, Jo, Kounovsky, Pablo, and Schwartz]{yu09}
Yu,~H.; Jo,~K.; Kounovsky,~K.~L.; Pablo,~J. J.~D.; Schwartz,~D.~C. {Molecular
  propulsion: Chemical sensing and chemotaxis of DNA driven by RNA polymerase}.
  \emph{J. Am. Chem. Soc.} \textbf{2009}, \emph{131}, 5722--5723\relax
\mciteBstWouldAddEndPuncttrue
\mciteSetBstMidEndSepPunct{\mcitedefaultmidpunct}
{\mcitedefaultendpunct}{\mcitedefaultseppunct}\relax
\EndOfBibitem
\bibitem[Muddana \latin{et~al.}(2010)Muddana, Sengupta, Mallouk, Sen, and
  Butler]{mudd10}
Muddana,~H.~S.; Sengupta,~S.; Mallouk,~T.~E.; Sen,~A.; Butler,~P.~J. {Substrate
  Catalysis Enhances Single Enzyme Diffusion}. \emph{J. Am. Chem. Soc.}
  \textbf{2010}, \emph{132}, 2110--2111\relax
\mciteBstWouldAddEndPuncttrue
\mciteSetBstMidEndSepPunct{\mcitedefaultmidpunct}
{\mcitedefaultendpunct}{\mcitedefaultseppunct}\relax
\EndOfBibitem
\bibitem[Sengupta \latin{et~al.}(2013)Sengupta, Dey, Muddana, Tabouillot,
  Ibele, Butler, and Sen]{seng13}
Sengupta,~S.; Dey,~K.~K.; Muddana,~H.~S.; Tabouillot,~T.; Ibele,~M.~E.;
  Butler,~P.~J.; Sen,~A. {Enzyme molecules as nanomotors}. \emph{J. Am. Chem.
  Soc.} \textbf{2013}, \emph{135}, 1406--1414\relax
\mciteBstWouldAddEndPuncttrue
\mciteSetBstMidEndSepPunct{\mcitedefaultmidpunct}
{\mcitedefaultendpunct}{\mcitedefaultseppunct}\relax
\EndOfBibitem
\bibitem[Sengupta \latin{et~al.}(2014)Sengupta, Spiering, Dey, Duan, Patra,
  Butler, Astumian, Benkovic, and Sen]{seng14}
Sengupta,~S.; Spiering,~M.~M.; Dey,~K.~K.; Duan,~W.; Patra,~D.; Butler,~P.~J.;
  Astumian,~R.~D.; Benkovic,~S.~J.; Sen,~A. {DNA polymerase as a molecular
  motor and pump}. \emph{ACS Nano} \textbf{2014}, \emph{8}, 2410--2418\relax
\mciteBstWouldAddEndPuncttrue
\mciteSetBstMidEndSepPunct{\mcitedefaultmidpunct}
{\mcitedefaultendpunct}{\mcitedefaultseppunct}\relax
\EndOfBibitem
\bibitem[Riedel \latin{et~al.}(2015)Riedel, Gabizon, Wilson, Hamadani,
  Tsekouras, Marqusee, Press{\'{e}}, and Bustamante]{ried15}
Riedel,~C.; Gabizon,~R.; Wilson,~C. A.~M.; Hamadani,~K.; Tsekouras,~K.;
  Marqusee,~S.; Press{\'{e}},~S.; Bustamante,~C. {The heat released during
  catalytic turnover enhances the diffusion of an enzyme.} \emph{Nature}
  \textbf{2015}, \emph{517}, 227--30\relax
\mciteBstWouldAddEndPuncttrue
\mciteSetBstMidEndSepPunct{\mcitedefaultmidpunct}
{\mcitedefaultendpunct}{\mcitedefaultseppunct}\relax
\EndOfBibitem
\bibitem[Illien \latin{et~al.}(2017)Illien, Zhao, Dey, Butler, Sen, and
  Golestanian]{illi17a}
Illien,~P.; Zhao,~X.; Dey,~K.~K.; Butler,~P.~J.; Sen,~A.; Golestanian,~R.
  {Exothermicity Is Not a Necessary Condition for Enhanced Diffusion of
  Enzymes}. \emph{Nano Lett.} \textbf{2017}, \emph{17}, 4415--4420\relax
\mciteBstWouldAddEndPuncttrue
\mciteSetBstMidEndSepPunct{\mcitedefaultmidpunct}
{\mcitedefaultendpunct}{\mcitedefaultseppunct}\relax
\EndOfBibitem
\bibitem[Golestanian(2010)]{gole10}
Golestanian,~R. {Synthetic mechanochemical molecular swimmer}. \emph{Phys. Rev.
  Lett.} \textbf{2010}, \emph{105}, 018103\relax
\mciteBstWouldAddEndPuncttrue
\mciteSetBstMidEndSepPunct{\mcitedefaultmidpunct}
{\mcitedefaultendpunct}{\mcitedefaultseppunct}\relax
\EndOfBibitem
\bibitem[Golestanian(2015)]{gole15}
Golestanian,~R. {Enhanced Diffusion of Enzymes that Catalyze Exothermic
  Reactions}. \emph{Phys. Rev. Lett.} \textbf{2015}, \emph{115}, 108102\relax
\mciteBstWouldAddEndPuncttrue
\mciteSetBstMidEndSepPunct{\mcitedefaultmidpunct}
{\mcitedefaultendpunct}{\mcitedefaultseppunct}\relax
\EndOfBibitem
\bibitem[Bai and Wolynes(2015)Bai, and Wolynes]{bai15}
Bai,~X.; Wolynes,~P.~G. {On the hydrodynamics of swimming enzymes}. \emph{J.
  Chem. Phys.} \textbf{2015}, \emph{143}, 165101\relax
\mciteBstWouldAddEndPuncttrue
\mciteSetBstMidEndSepPunct{\mcitedefaultmidpunct}
{\mcitedefaultendpunct}{\mcitedefaultseppunct}\relax
\EndOfBibitem
\bibitem[Hwang and Hyeon(2017)Hwang, and Hyeon]{hwan17}
Hwang,~W.; Hyeon,~C. {Quantifying the Heat Dissipation from a Molecular Motor's
  Transport Properties in Nonequilibrium Steady States}. \emph{J. Phys. Chem.
  Lett.} \textbf{2017}, \emph{8}, 250--256\relax
\mciteBstWouldAddEndPuncttrue
\mciteSetBstMidEndSepPunct{\mcitedefaultmidpunct}
{\mcitedefaultendpunct}{\mcitedefaultseppunct}\relax
\EndOfBibitem
\bibitem[Illien \latin{et~al.}(2017)Illien, Adeleke-Larodo, and
  Golestanian]{illi17b}
Illien,~P.; Adeleke-Larodo,~T.; Golestanian,~R. {Diffusion of an enzyme: The
  role of fluctuation-induced hydrodynamic coupling}. \emph{EPL} \textbf{2017},
  \emph{119}, 40002\relax
\mciteBstWouldAddEndPuncttrue
\mciteSetBstMidEndSepPunct{\mcitedefaultmidpunct}
{\mcitedefaultendpunct}{\mcitedefaultseppunct}\relax
\EndOfBibitem
\bibitem[Dey \latin{et~al.}(2014)Dey, Das, Poyton, Sengupta, Butler, Cremer,
  and Sen]{dey14}
Dey,~K.~K.; Das,~S.; Poyton,~M.~F.; Sengupta,~S.; Butler,~P.~J.; Cremer,~P.~S.;
  Sen,~A. {Chemotactic separation of enzymes}. \emph{ACS Nano} \textbf{2014},
  \emph{8}, 11941--11949\relax
\mciteBstWouldAddEndPuncttrue
\mciteSetBstMidEndSepPunct{\mcitedefaultmidpunct}
{\mcitedefaultendpunct}{\mcitedefaultseppunct}\relax
\EndOfBibitem
\bibitem[Zhao \latin{et~al.}(2018)Zhao, Palacci, Yadav, Spiering, Gilson,
  Butler, Hess, Benkovic, and Sen]{zhao17}
Zhao,~X.; Palacci,~H.; Yadav,~V.; Spiering,~M.~M.; Gilson,~M.~K.;
  Butler,~P.~J.; Hess,~H.; Benkovic,~S.~J.; Sen,~A. {Substrate-driven
  chemotactic assembly in an enzyme cascade}. \emph{Nat. Chem.} \textbf{2018},
  \emph{10}, 311--317\relax
\mciteBstWouldAddEndPuncttrue
\mciteSetBstMidEndSepPunct{\mcitedefaultmidpunct}
{\mcitedefaultendpunct}{\mcitedefaultseppunct}\relax
\EndOfBibitem
\bibitem[Jee \latin{et~al.}(2018)Jee, Dutta, Cho, Tlusty, and Granick]{jee17}
Jee,~A.-Y.; Dutta,~S.; Cho,~Y.-K.; Tlusty,~T.; Granick,~S. {Enzyme leaps fuel
  antichemotaxis}. \emph{Proc. Natl. Acad. Sci. U. S. A.} \textbf{2018},
  \emph{115}, 14--18\relax
\mciteBstWouldAddEndPuncttrue
\mciteSetBstMidEndSepPunct{\mcitedefaultmidpunct}
{\mcitedefaultendpunct}{\mcitedefaultseppunct}\relax
\EndOfBibitem
\bibitem[Guha \latin{et~al.}(2017)Guha, Mohajerani, Collins, Ghosh, Sen, and
  Velegol]{guha17}
Guha,~R.; Mohajerani,~F.; Collins,~M.; Ghosh,~S.; Sen,~A.; Velegol,~D.
  {Chemotaxis of Molecular Dyes in Polymer Gradients in Solution}. \emph{J. Am.
  Chem. Soc.} \textbf{2017}, \emph{139}, 15588--15591\relax
\mciteBstWouldAddEndPuncttrue
\mciteSetBstMidEndSepPunct{\mcitedefaultmidpunct}
{\mcitedefaultendpunct}{\mcitedefaultseppunct}\relax
\EndOfBibitem
\bibitem[Lee and Karplus(1987)Lee, and Karplus]{lee87}
Lee,~S.; Karplus,~M. {Kinetics of diffusion-influenced bimolecular reactions in
  solution. I. General formalism and relaxation kinetics of fast reversible
  reactions}. \emph{J. Chem. Phys.} \textbf{1987}, \emph{86}, 1883--1903\relax
\mciteBstWouldAddEndPuncttrue
\mciteSetBstMidEndSepPunct{\mcitedefaultmidpunct}
{\mcitedefaultendpunct}{\mcitedefaultseppunct}\relax
\EndOfBibitem
\bibitem[Derjaguin \latin{et~al.}(1947)Derjaguin, Sidorenkov, Zubashchenkov,
  and Kiseleva]{derj47}
Derjaguin,~B.~V.; Sidorenkov,~G.~P.; Zubashchenkov,~E.~A.; Kiseleva,~E.~V.
  {Kinetic phenomena in boundary films of liquids}. \emph{Kolloidn. Zh}
  \textbf{1947}, \emph{9}, 335--47\relax
\mciteBstWouldAddEndPuncttrue
\mciteSetBstMidEndSepPunct{\mcitedefaultmidpunct}
{\mcitedefaultendpunct}{\mcitedefaultseppunct}\relax
\EndOfBibitem
\bibitem[Ebbens \latin{et~al.}(2012)Ebbens, Tu, Howse, and Golestanian]{ebbe12}
Ebbens,~S.; Tu,~M.~H.; Howse,~J.~R.; Golestanian,~R. {Size dependence of the
  propulsion velocity for catalytic Janus-sphere swimmers}. \emph{Phys. Rev. E}
  \textbf{2012}, \emph{85}, 020401(R)\relax
\mciteBstWouldAddEndPuncttrue
\mciteSetBstMidEndSepPunct{\mcitedefaultmidpunct}
{\mcitedefaultendpunct}{\mcitedefaultseppunct}\relax
\EndOfBibitem
\bibitem[Anderson(1989)]{ande89}
Anderson,~J.~L. {Colloid transport by interfacial forces}. \emph{Annu. Rev.
  Fluid Mech.} \textbf{1989}, \emph{21}, 61--99\relax
\mciteBstWouldAddEndPuncttrue
\mciteSetBstMidEndSepPunct{\mcitedefaultmidpunct}
{\mcitedefaultendpunct}{\mcitedefaultseppunct}\relax
\EndOfBibitem
\bibitem[Weistuch and Press{\'{e}}(2017)Weistuch, and Press{\'{e}}]{weis17}
Weistuch,~C.; Press{\'{e}},~S. {Spatiotemporal Organization of Catalysts Driven
  by Enhanced Diffusion}. \emph{J. Phys. Chem. B} \textbf{2017},
  doi:acs.jpcb.7b06868\relax
\mciteBstWouldAddEndPuncttrue
\mciteSetBstMidEndSepPunct{\mcitedefaultmidpunct}
{\mcitedefaultendpunct}{\mcitedefaultseppunct}\relax
\EndOfBibitem
\bibitem[Schnitzer(1993)]{schn93}
Schnitzer,~M.~J. {Theory of continuum random walks and application to
  chemotaxis}. \emph{Phys. Rev. E} \textbf{1993}, \emph{48}, 2553--2568\relax
\mciteBstWouldAddEndPuncttrue
\mciteSetBstMidEndSepPunct{\mcitedefaultmidpunct}
{\mcitedefaultendpunct}{\mcitedefaultseppunct}\relax
\EndOfBibitem
\bibitem[Lau and Lubensky(2007)Lau, and Lubensky]{lau07}
Lau,~A. W.~C.; Lubensky,~T.~C. {State-dependent diffusion: Thermodynamic
  consistency and its path integral formulation}. \emph{Phys. Rev. E}
  \textbf{2007}, \emph{76}, 011123\relax
\mciteBstWouldAddEndPuncttrue
\mciteSetBstMidEndSepPunct{\mcitedefaultmidpunct}
{\mcitedefaultendpunct}{\mcitedefaultseppunct}\relax
\EndOfBibitem
\bibitem[Krajewska(2009)]{kraj09}
Krajewska,~B. {Ureases I. Functional, catalytic and kinetic properties: A
  review}. \emph{J. Mol. Catal. B Enzym.} \textbf{2009}, \emph{59}, 9--21\relax
\mciteBstWouldAddEndPuncttrue
\mciteSetBstMidEndSepPunct{\mcitedefaultmidpunct}
{\mcitedefaultendpunct}{\mcitedefaultseppunct}\relax
\EndOfBibitem
\bibitem[Switala and Loewen(2002)Switala, and Loewen]{swit02}
Switala,~J.; Loewen,~P.~C. {Diversity of properties among catalases}.
  \emph{Arch. Biochem. Biophys.} \textbf{2002}, \emph{401}, 145--154\relax
\mciteBstWouldAddEndPuncttrue
\mciteSetBstMidEndSepPunct{\mcitedefaultmidpunct}
{\mcitedefaultendpunct}{\mcitedefaultseppunct}\relax
\EndOfBibitem
\bibitem[Hu \latin{et~al.}(2006)Hu, Jiang, Xu, Pan, and Zou]{hu06}
Hu,~L.; Jiang,~G.; Xu,~S.; Pan,~C.; Zou,~H. {Monitoring Enzyme Reaction and
  Screening Enzyme Inhibitor Based on MALDI-TOF-MS Platform with a Matrix of
  Oxidized Carbon Nanotubes}. \emph{J. Am. Soc. Mass Spectrom.} \textbf{2006},
  \emph{17}, 1616--1619\relax
\mciteBstWouldAddEndPuncttrue
\mciteSetBstMidEndSepPunct{\mcitedefaultmidpunct}
{\mcitedefaultendpunct}{\mcitedefaultseppunct}\relax
\EndOfBibitem
\bibitem[Pohanka \latin{et~al.}(2011)Pohanka, Hrabinova, Kuca, and
  Simonato]{poha11}
Pohanka,~M.; Hrabinova,~M.; Kuca,~K.; Simonato,~J.~P. {Assessment of
  acetylcholinesterase activity using indoxylacetate and comparison with the
  standard Ellman's method}. \emph{Int. J. Mol. Sci.} \textbf{2011}, \emph{12},
  2631--2640\relax
\mciteBstWouldAddEndPuncttrue
\mciteSetBstMidEndSepPunct{\mcitedefaultmidpunct}
{\mcitedefaultendpunct}{\mcitedefaultseppunct}\relax
\EndOfBibitem
\bibitem[Saha \latin{et~al.}(2014)Saha, Golestanian, and Ramaswamy]{saha14}
Saha,~S.; Golestanian,~R.; Ramaswamy,~S. {Clusters, asters, and collective
  oscillations in chemotactic colloids}. \emph{Phys. Rev. E} \textbf{2014},
  \emph{89}, 062316\relax
\mciteBstWouldAddEndPuncttrue
\mciteSetBstMidEndSepPunct{\mcitedefaultmidpunct}
{\mcitedefaultendpunct}{\mcitedefaultseppunct}\relax
\EndOfBibitem
\bibitem[Mikhailov and Kapral(2015)Mikhailov, and Kapral]{mikh15}
Mikhailov,~A.~S.; Kapral,~R. {Hydrodynamic collective effects of active protein
  machines in solution and lipid bilayers.} \emph{Proc. Natl. Acad. Sci. U. S.
  A.} \textbf{2015}, \emph{112}, E3639--3644\relax
\mciteBstWouldAddEndPuncttrue
\mciteSetBstMidEndSepPunct{\mcitedefaultmidpunct}
{\mcitedefaultendpunct}{\mcitedefaultseppunct}\relax
\EndOfBibitem
\bibitem[Felderhof(1977)]{feld77}
Felderhof,~B.~U. {Hydrodynamic interaction between two spheres}. \emph{Physica
  A} \textbf{1977}, \emph{89}, 373--384\relax
\mciteBstWouldAddEndPuncttrue
\mciteSetBstMidEndSepPunct{\mcitedefaultmidpunct}
{\mcitedefaultendpunct}{\mcitedefaultseppunct}\relax
\EndOfBibitem
\end{mcitethebibliography}


\providecommand{\latin}[1]{#1}
\makeatletter
\providecommand{\doi}
  {\begingroup\let\do\@makeother\dospecials
  \catcode`\{=1 \catcode`\}=2 \doi@aux}
\providecommand{\doi@aux}[1]{\endgroup\texttt{#1}}
\makeatother
\providecommand*\mcitethebibliography{\thebibliography}
\csname @ifundefined\endcsname{endmcitethebibliography}
  {\let\endmcitethebibliography\endthebibliography}{}
\begin{mcitethebibliography}{3}
\providecommand*\natexlab[1]{#1}
\providecommand*\mciteSetBstSublistMode[1]{}
\providecommand*\mciteSetBstMaxWidthForm[2]{}
\providecommand*\mciteBstWouldAddEndPuncttrue
  {\def\EndOfBibitem{\unskip.}}
\providecommand*\mciteBstWouldAddEndPunctfalse
  {\let\EndOfBibitem\relax}
\providecommand*\mciteSetBstMidEndSepPunct[3]{}
\providecommand*\mciteSetBstSublistLabelBeginEnd[3]{}
\providecommand*\EndOfBibitem{}
\mciteSetBstSublistMode{f}
\mciteSetBstMaxWidthForm{subitem}{(\alph{mcitesubitemcount})}
\mciteSetBstSublistLabelBeginEnd
  {\mcitemaxwidthsubitemform\space}
  {\relax}
  {\relax}

\bibitem[Felderhof(1977)]{feld77}
Felderhof,~B.~U. {Hydrodynamic interaction between two spheres}. \emph{Physica
  A} \textbf{1977}, \emph{89}, 373--384\relax
\mciteBstWouldAddEndPuncttrue
\mciteSetBstMidEndSepPunct{\mcitedefaultmidpunct}
{\mcitedefaultendpunct}{\mcitedefaultseppunct}\relax
\EndOfBibitem
\bibitem[Anderson(1989)]{ande89}
Anderson,~J.~L. {Colloid transport by interfacial forces}. \emph{Annu. Rev.
  Fluid Mech.} \textbf{1989}, \emph{21}, 61--99\relax
\mciteBstWouldAddEndPuncttrue
\mciteSetBstMidEndSepPunct{\mcitedefaultmidpunct}
{\mcitedefaultendpunct}{\mcitedefaultseppunct}\relax
\EndOfBibitem
\end{mcitethebibliography}

\end{document}





\setcounter{page}{1}
\renewcommand{\thepage}{S\arabic{page}}

\renewcommand{\thefigure}{S\arabic{figure}}
\setcounter{figure}{0}

\section{Obtaining the one-particle distribution functions}

Integrating Equations~(1--2) in the main text with respect to all the position coordinates of the free substrate molecules, we obtain
\begin{eqnarray}
\partial_t \rho_{\ee}(\RR;t) & = & \nabla_{\RR}\cdot \left[  \kB T \boldsymbol{\mu}^{\ee \ee} \nabla_{\RR}\rho_{\ee}   \right] +  N \nabla_{\RR}  \cdot  \int_{\XX} (\boldsymbol{\mu}^{\ee \sss} - \boldsymbol{\mu}^{\ee \ee}) [\kB T \nabla_{\XX} \rho_{\ee \sss}^{(2)} + (\nabla_{\XX} \phi^{\ee \sss}) \rho_{\ee \sss}^{(2)}  ]   \nonumber\\
&&- N  \int_{\XX} S_\text{on}(\RR,\XX) \rho_{\ee \sss}^{(2)}  + \rho_{\cc} \int_{\XX} S_\text{off}(\RR,\XX) \label{eq:twocoup1} \\
\partial_t \rho_{\cc,i}(\RR;t) & = & \nabla_{\RR}\cdot \left[  \kB T \boldsymbol{\mu}^{\cc \cc} \nabla_{\RR}\rho_{\cc,i}   \right] +  (N-1) \nabla_{\RR}  \cdot  \int_{\XX} (\boldsymbol{\mu}^{\cc \sss} - \boldsymbol{\mu}^{\cc \cc}) [\kB T \nabla_{\XX} \rho_{\cc \sss,i}^{(2)} + (\nabla_{\XX} \phi^{\cc \sss}) \rho_{\cc \sss,i}^{(2)}  ]   \nonumber\\
&&+  \int_{\XX} S_\text{on}(\RR,\XX) \rho_{\ee \sss}^{(2)}  - \rho_{\cc,i} \int_{\XX} S_\text{off}(\RR,\XX) \label{eq:twocoup2}
\end{eqnarray}
which is not a closed equation, because it depends on the two-particle probability distributions $\rho_{\ee \sss}^{(2)}(\RR,\XX)$ and $\rho_{\cc \sss,i}^{(2)}(\RR,\XX)$. If we tried to obtain an evolution equation for these two-particle distributions, the latter would depend on the three-particle distibutions, and so on, leading to an infinite hierarchy of equations of the BBGKY type. In order to close the hierarchy, we use the following approximate forms for the two-point distribution functions
\begin{eqnarray}
\rho_{\ee \sss}^{(2)}(\RR,\XX;t) & = & \rho_{\ee}(\RR;t) \frac{\rho_{\sss}(\XX)}{N}\ex{- \phi^{\ee \sss}(\RR-\XX)/\kB T}, \\
\rho_{\cc \sss,i}^{(2)}(\RR,\XX;t) & = & \rho_{\cc,i}(\RR;t)\frac{\rho_{\sss}(\XX)}{N-1}\ex{- \phi^{\cc \sss}(\RR-\XX)/\kB T}. 
\end{eqnarray}
where the $N$ and $N-1$ denominators ensure that $\rho_{\sss}$ corresponds to the concentration of substrate molecules. Substituting these expressions into (\ref{eq:twocoup1}--\ref{eq:twocoup2}), we finally obtain the closed-form coupled system of equations
\begin{eqnarray}
\partial_t \rho_{\ee}(\RR;t) & = &  \nabla_{\RR}\cdot \left[  D_{\ee} \nabla_{\RR}\rho_{\ee}  - \boldsymbol{v_{\ee}}(\RR) \rho_{\ee} \right]  - \mu(\RR) \rho_{\ee} + \nu(\RR) \rho_{\cc}  \\
\partial_t \rho_{\cc}(\RR;t) & = &  \nabla_{\RR}\cdot \left[  D_{\cc} \nabla_{\RR}\rho_{\cc}  - \boldsymbol{v_{\cc}}(\RR) \rho_{\cc} \right]  + \mu(\RR) \rho_{\ee} - \nu(\RR) \rho_{\cc}
\end{eqnarray}
with the diffusion coefficients, drift velocities and reaction rates
\begin{eqnarray}
D_{\ee} & \equiv & \kB T \boldsymbol{\mu}^{\ee \ee} \\
D_{\cc} & \equiv & \kB T \boldsymbol{\mu}^{\cc \cc} \\
\boldsymbol{v_{\ee}}(\RR) & \equiv & - \kB T   \int_{\XX} (\boldsymbol{\mu}^{\ee \sss} - \boldsymbol{\mu}^{\ee \ee})   ( \ex{- \phi^{\ee \sss}(\RR-\XX)/\kB T} - 1  ) \nabla_{\XX} \rho_{\sss} \label{eq:vA} \\
\boldsymbol{v_{\cc}}(\RR) & \equiv & - \kB T   \int_{\XX} (\boldsymbol{\mu}^{\cc \sss} - \boldsymbol{\mu}^{\cc \cc})   ( \ex{- \phi^{\cc \sss}(\RR-\XX)/\kB T} - 1  ) \nabla_{\XX} \rho_{\sss} \label{eq:vB} \\
\mu(\RR) & \equiv & \int_{\XX} S_\text{on}(\RR,\XX) \rho_{\sss} \ex{- \phi^{\ee \sss}(\RR-\XX)/\kB T} \\
\nu(\RR) & \equiv &  \int_{\XX} S_\text{off}(\RR,\XX)
\end{eqnarray}
Throughout this derivation, we have used the fact that the mobilities are divergence free, i.e.~$\nabla_{\XX} \boldsymbol{\mu}^{\ee \ee} = \nabla_{\XX} \boldsymbol{\mu}^{\cc \cc} = \nabla_{\XX} \boldsymbol{\mu}^{\ee \sss} = \nabla_{\XX} \boldsymbol{\mu}^{\cc \sss} = 0$. This is exactly true if the substrate particles are point-like, and is also true for the mobility expanded up to order $O(1/r^6)$ in the distance between particles if substrate particles are not point-like. \cite{feld77} Furthermore, we have ignored boundary terms in the integrals, and we have made use of the identities $\nabla_{\RR} \phi^{\ee \sss} = - \nabla_{\XX} \phi^{\ee \sss}$ and $\nabla_{\RR} \phi^{\cc \sss} = - \nabla_{\XX} \phi^{\cc \sss}$ for the pair potentials.
 
As described in the main text, for the simplest description of binding, we may use $S_\text{on}(\RR,\XX) = k_\text{on} \delta(\RR-\XX)$ and $S_\text{off}(\RR,\XX) = k_\text{off} \delta(\RR-\XX)$. In this case, we find $\mu(\RR) = k_\mathrm{on} \ex{- \phi^{\ee \sss}(0)/\kB T} \rho_{\sss}(\RR)$ and $\nu(\RR) = k_\mathrm{off}$. Because $\ex{- \phi^{\ee \sss}(0)/\kB T}$ is simply a constant, we can absorb it by definition into $k_\text{on}$ and simply write $\mu(\RR) = k_\mathrm{on} \rho_{\sss}(\RR)$, without loss of generality, as has been done in the main text. The expressions for the drift velocities $\boldsymbol{v_{\ee}}(\RR)$ and $\boldsymbol{v_{\cc}}(\RR)$ in (\ref{eq:vA}--\ref{eq:vB}) can also be further simplified, as described in the next section.

\section{Diffusiophoretic velocities $\boldsymbol{v_{\ee}}(\RR)$ and $\boldsymbol{v_{\cc}}(\RR)$}

We show here how the expressions (\ref{eq:vA}--\ref{eq:vB}) for the diffusiophoretic velocities can be further simplified. We first note that the expressions for the free enzyme ($\boldsymbol{v_{\ee}}$) and the complex ($\boldsymbol{v_{\cc}}$) are completely analogous. We therefore describe here the derivation only for the free enzyme, the result for the complex follows by simply replacing all the labels `e' with `c'.

For a spherical enzyme of radius $R_\mathrm{e}$, assuming the substrate molecules are point-like (i.e.~significantly smaller), we have the hydrodynamic mobility tensors
\begin{equation}
\boldsymbol{\mu}^{\ee \ee} = \frac{1}{6 \pi \eta R_\mathrm{e}} \boldsymbol{I}
\end{equation}
\begin{equation}
\boldsymbol{\mu}^{\ee \sss} = \frac{1}{6 \pi \eta R_\mathrm{e}} \left[ \frac{1}{4} \left( 3 \frac{R_\mathrm{e}}{r} + \frac{R_\mathrm{e}^3}{r^3} \right) \boldsymbol{I} + \frac{3}{4} \left(\frac{R_\mathrm{e}}{r} - \frac{R_\mathrm{e}^3}{r^3} \right) \boldsymbol{\hat{r}}\boldsymbol{\hat{r}} \right]
\end{equation}
where $r$ is the distance between the center of the enzyme and the substrate molecule, and $\boldsymbol{\hat{r}}$ is the radial unit vector. Combining both, we have
\begin{equation}
\boldsymbol{\mu}^{\ee \sss} - \boldsymbol{\mu}^{\ee \ee} = \frac{1}{6 \pi \eta R_\mathrm{e}} \left[  \left( -1 + \frac{3}{4} \frac{R_\mathrm{e}}{r} + \frac{1}{4} \frac{R_\mathrm{e}^3}{r^3} \right) (\boldsymbol{I}-\boldsymbol{\hat{r}}\boldsymbol{\hat{r}}) + \left(-1 + \frac{3}{2} \frac{R_\mathrm{e}}{r} - \frac{1}{2}\frac{R_\mathrm{e}^3}{r^3} \right) \boldsymbol{\hat{r}}\boldsymbol{\hat{r}} \right]
\end{equation}

The concentration of substrate molecules in the proximity of the enzyme, assuming that far enough from the enzyme there is a concentration gradient $\nabla \rho_{\sss}^\infty = |\nabla \rho_{\sss}^\infty| \boldsymbol{\hat{z}}$ of substrate molecules pointing in the $z$-direction, is given by the solution to the Laplace equation with no normal flux boundary conditions on the surface of the sphere
\begin{equation}
\rho_{\sss} (r,\theta,\varphi) = A + |\nabla \rho_{\sss}^\infty| \left( r + \frac{1}{2} \frac{R_\mathrm{e}^3}{r^2} \right) \cos \theta
\end{equation}
where we use spherical coordinates centred on the enzyme. Using the gradient operator in spherical coordinates $\nabla f = \partial_r f \boldsymbol{\hat{r}} + (1/r) \partial_\theta f \boldsymbol{\hat{\theta}} + (1/r \sin \theta) \partial_\varphi f \boldsymbol{\hat{\varphi}}$, we calculate
\begin{equation}
\nabla \rho_{\sss}  =  |\nabla \rho_{\sss}^\infty| \left[ \left( 1 - \frac{R_\mathrm{e}^3}{r^3} \right) \cos \theta \boldsymbol{\hat{r}} - \left( 1 + \frac{1}{2} \frac{R_\mathrm{e}^3}{r^3} \right) \sin \theta \boldsymbol{\hat{\theta}}   \right]
\end{equation}

In order to evaluate (\ref{eq:vA}), we need to calculate $(\boldsymbol{I}-\boldsymbol{\hat{r}}\boldsymbol{\hat{r}})\nabla \rho_{\sss}$ and $\boldsymbol{\hat{r}}\boldsymbol{\hat{r}} \nabla \rho_{\sss}$. Using $(\boldsymbol{I}-\boldsymbol{\hat{r}}\boldsymbol{\hat{r}}) \boldsymbol{\hat{r}} = 0$, $(\boldsymbol{I}-\boldsymbol{\hat{r}}\boldsymbol{\hat{r}}) \boldsymbol{\hat{\theta}} = \boldsymbol{\hat{\theta}}$, $(\boldsymbol{\hat{r}}\boldsymbol{\hat{r}})\boldsymbol{\hat{r}} = \boldsymbol{\hat{r}}$, and $(\boldsymbol{\hat{r}}\boldsymbol{\hat{r}}) \boldsymbol{\hat{\theta}}=0$, we obtain
\begin{equation}
(\boldsymbol{I}-\boldsymbol{\hat{r}}\boldsymbol{\hat{r}}) \nabla \rho_{\sss} = - |\nabla \rho_{\sss}^\infty| \left( 1 + \frac{1}{2} \frac{R_\mathrm{e}^3}{r^3} \right) \sin \theta \boldsymbol{\hat{\theta}}
\end{equation}
\begin{equation}
(\boldsymbol{\hat{r}}\boldsymbol{\hat{r}}) \nabla \rho_{\sss} = |\nabla \rho_{\sss}^\infty|  \left( 1 - \frac{R_\mathrm{e}^3}{r^3} \right) \cos \theta \boldsymbol{\hat{r}}
\end{equation}
Furthermore, using the definition in cartesian coordinates of $\boldsymbol{\hat{r}} = \sin \theta \cos \varphi \boldsymbol{\hat{x}} + \sin \theta \sin \varphi \boldsymbol{\hat{y}} + \cos \theta  \boldsymbol{\hat{z}}$ and $\boldsymbol{\hat{\theta}} = \cos \theta \cos \varphi \boldsymbol{\hat{x}} + \cos \theta \sin \varphi \boldsymbol{\hat{y}} - \sin \theta  \boldsymbol{\hat{z}}$, we can calculate the integrals over the solid angle $\int \mathrm{d}\Omega \sin \theta \boldsymbol{\hat{\theta}} = -(8\pi/3) \boldsymbol{\hat{z}}$ and $\int \mathrm{d}\Omega \cos \theta \boldsymbol{\hat{r}} = (4\pi/3) \boldsymbol{\hat{z}}$.

Taking together all these results, and noting that the two sets of coordinates used are related to each other by $\XX = \RR + r \boldsymbol{\hat{r}}$, we can finally evaluate  (\ref{eq:vA}) to be
\begin{equation}
\boldsymbol{v_{\ee}}(\RR) = - \frac{\kB T}{6 \eta R_\mathrm{e}} \nabla_{\RR} \rho_{\sss}^\infty(\RR) \int_{R_\mathrm{e}}^\infty \mathrm{d}r r^2 (1 - \ex{- \phi^{\ee \sss}(r)/\kB T} ) \left( 4 - 4 \frac{R_\mathrm{e}}{r} + \frac{R_\mathrm{e}^4}{r^4} - \frac{R_\mathrm{e}^6}{r^6} \right)
\label{eq:anderson}
\end{equation}
This expression is a generalization of the classic formula for the diffusiophoretic velocity \cite{ande89} to cases in which the range of the interactions is not necessarily much smaller than the particle radius. Indeed, for very short range interactions we can use $r = R_\mathrm{e} + h$ with $h \ll a$. The terms inside the rightmost parenthesis in the integral become $6 h/R_\mathrm{e}$ to lowest order, giving back the classic formula
\begin{equation}
\boldsymbol{v_{\ee}}(\RR) \simeq \frac{\kB T}{\eta} \nabla_{\RR} \rho_{\sss}^\infty(\RR) \int_0^\infty \mathrm{d}h h (\ex{- \phi^{\ee \sss}(h)/\kB T} - 1) \equiv \frac{\kB T}{\eta} \lambda_{\ee}^2 \nabla_{\RR} \rho_{\sss}^\infty
\end{equation}
which is the expression for the diffusiophoretic velocity in Equation (8) of the main text. Because we are ultimately interested only in the `coarse-grained', mesoscale concentration field of substrate molecules in lengthscales much larger than the size of the enzyme that is accessible to experiment, we have omitted the superscript of $\rho_{\sss}^\infty$ in the main text.


\section{Estimating the Derjaguin length for an attractive potential}

\begin{figure}
\begin{center}
\includegraphics[width=0.65\linewidth]{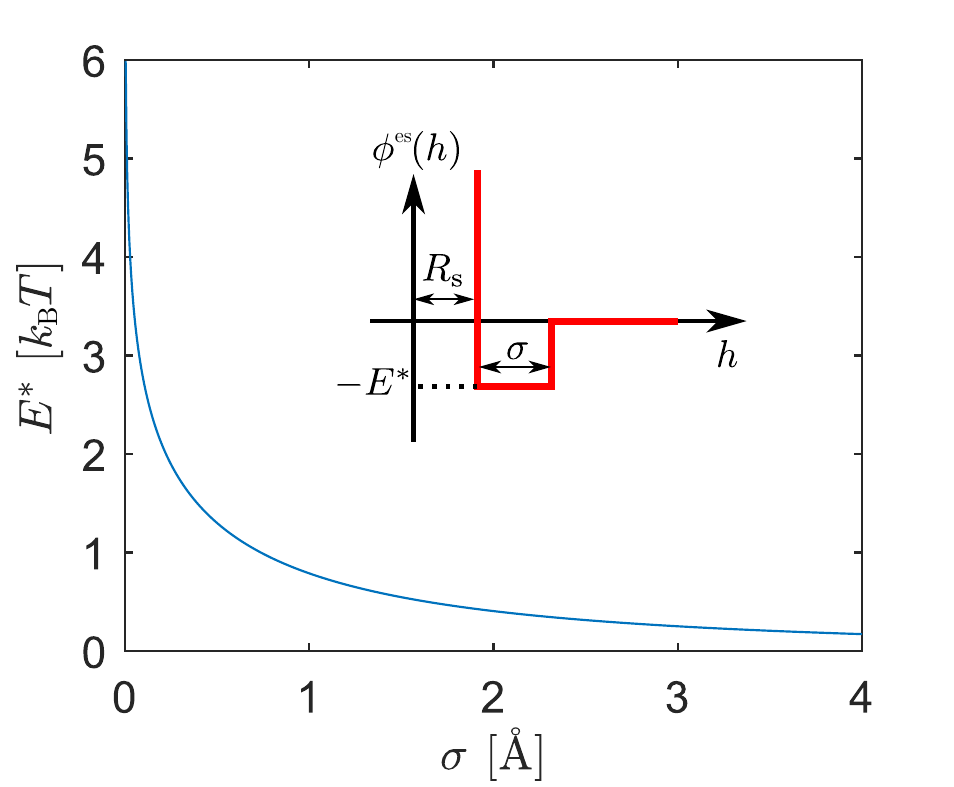}
\caption{Depth $E^*$ of the attractive interaction potential necessary to obtain attractive phoresis with $\lambda_{\ee}^2>0$ and $\lambda_{\ee}=1$~\AA, for hard substrate molecules with radius $R_\mathrm{s}=2$~\AA, as a function of the potential range $\sigma$.}
\label{fig:squarepotential}
\end{center}
\end{figure}

We consider an attractive interaction potential $\phi^{\ee \sss}(h)$ between enzyme and substrate, where $h$ is the distance between the surface of the enzyme and the center of the substrate molecule. In order to get a clear insight into the respective effects of the depth and the range of the potential, we consider a square-well potential such as the one in the inset of Fig.~\ref{fig:squarepotential}, with depth $E^*$ and range $\sigma$. The infinite wall at $h=R_\mathrm{s}$ represents the hard-sphere repulsion between the enzyme and the substrate, where $R_{\sss}$ is the radius of the substrate molecule.

For such a potential, the Derjaguin length can be directly calculated as
\begin{equation}
\lambda_{\ee}^2 = \int_0^\infty \mathrm{d}h h \left( \ex{- \phi^{\ee \sss}(h)/\kB T} -1 \right) = -\frac{R_\mathrm{s}^2}{2} + \frac{\ex{E^*/\kB T} -1}{2} [(R_\mathrm{s}+\sigma)^2-R_\mathrm{s}^2]
\label{eq:derjsq}
\end{equation}
Here, the first term arises from the hard-sphere repulsion between enzyme and substrate,\cite{ande89} and the second term from the attractive part of the potential.

Because the hard-sphere repulsion leads to a negative contribution to (\ref{eq:derjsq}), in order to obtain an overall positive $\lambda_{\ee}^2$ we need a sufficiently deep or sufficiently long-ranged attractive part. But how deep or how long-ranged exactly? Rearranging the equation above, we obtain
\begin{equation}
\frac{E^*}{\kB T} = \log \left( 1 + \frac{2 \lambda_{\ee}^2 + R_\mathrm{s}^2}{\sigma (\sigma + 2 R_\mathrm{s})} \right)
\end{equation}
which gives us the necessary potential depth $E^*$ for given Derjaguin length $\lambda_{\ee}$, potential range $\sigma$, and substrate radius $R_{\sss}$. Let us consider the case of urease--urea, discussed in the main text, as an example. Using the hydrodynamic radius of urea $R_\mathrm{s}=2$~\AA, we can calculate the potential depth required to achieve a positive value of $\lambda_{\ee}^2>0$ with $\lambda_{\ee}=1$~\AA, as a function of the potential range. This is plotted in Fig.~\ref{fig:squarepotential}. We find that a rather weak ($E^* = 1~\kB T$) and short-ranged ($\sigma = 0.75$~\AA) attractive interaction is already sufficient to counteract the steric repulsion and induce attractive phoresis with $\lambda_{\ee}^2>0$, $\lambda_{\ee}=1$~\AA. Such a weak and short ranged attractive interaction could be provided, for example, by van der Waals forces.

We note that the short range of the potential obtained in this way (a few angstroms at most) self-consistently validates the assumption of potential ranges much smaller than the enzyme size used in deriving the simplfied expression $\lambda_{\ee}^2 = \int_0^\infty \mathrm{d}h h \left( \ex{- \phi^{\ee \sss}(h)/\kB T} -1 \right)$ for the Derjaguin length. In this limit, the Derjaguin length is independent of the enzyme radius.

\bibliography{biblio}